\documentclass[
prl,
twocolumn,
superscriptaddress,
a4paper,
showpacs,
showkeys,
floatfix
]{revtex4}
\usepackage[final]{graphicx}
\usepackage{amsmath}
\usepackage{amssymb}

\newcommand{\mrm}[1]{\mathrm{#1}}

\newcommand{\xn}{x_\mrm{N}}

\newcommand{\istar}{i^*}

\newcommand{\vect}[1]{%
  \mathbf{#1}
}

\newcommand{\dimless}[1]{%
  \tilde{#1}
}

\begin{document}

\title{Dynamic effects on the loss of control in template-directed nucleation}

\date{\today}

\author{Felix Kalischewski}
\affiliation{\frenchspacing Westf\"alische Wilhelms Universit\"at M\"unster, Institut f\"ur physikalische Chemie, Corrensstr.\ 30, 48149 M\"unster, Germany}
\affiliation{\frenchspacing Center of Nonlinear Science CeNoS, Westf\"alische Wilhelms Universit\"at M\"unster, Germany}
\affiliation{\frenchspacing NRW Graduate School of Chemistry, Westf\"alische Wilhelms Universit\"at M\"unster, Germany}
\author{Andreas Heuer}
\affiliation{\frenchspacing Westf\"alische Wilhelms Universit\"at M\"unster, Institut f\"ur physikalische Chemie, Corrensstr.\ 30, 48149 M\"unster, Germany}
\affiliation{\frenchspacing Center of Nonlinear Science CeNoS, Westf\"alische Wilhelms Universit\"at M\"unster, Germany}
\affiliation{\frenchspacing NRW Graduate School of Chemistry, Westf\"alische Wilhelms Universit\"at M\"unster, Germany}

\begin{abstract}
Full nucleation control for deposited functional molecules on pre-patterned surfaces is of major technological relevance. To understand the nucleation behavior we combine the numerical solution for the evolution of the adatom concentration with standard nucleation theory. From the qualitative change in nucleation behavior upon variation of the pattern spacing and coverage we show why the quality of nucleation control can vary significantly in different parameter regimes. In some limits analytical expressions can be formulated for the nucleation control. Our analysis provides a theoretical explanation for previous experimental observations [Wang et al, PRL 98, 225504 (2007)].

\end{abstract}

\pacs{81.65.Cf}

\keywords{Nucleation control, template-directed nucleation}

\maketitle

The controlled fabrication of nano-structures is a technologically
highly demanding and interesting topic. In this context techniques
based on lithography and etching have become very sophisticated
over the past decades. However, with the advent of functional
organic molecules in micro-electronics, one faces the problem that
this technique is often not applicable to this class of
substances. At this point the concept of template-directed growth
becomes interesting: in this technique the areas of desired
accretion are marked in some fashion to favor the adsorption of
the functional substance, which is subsequently deposited. There
exists a number of methods to generate these preferred sites; see,
e.g.,
Refs.~\cite{BardottiNanoclusters,LaserAblation,NatureDirectImprinting,AFMPatterningHyon,AFMPatterning,BrisenoNature,DipPenWritingLehnert,ChiPRL}.

\begin{figure}[ht]
\includegraphics[width=0.95\linewidth]{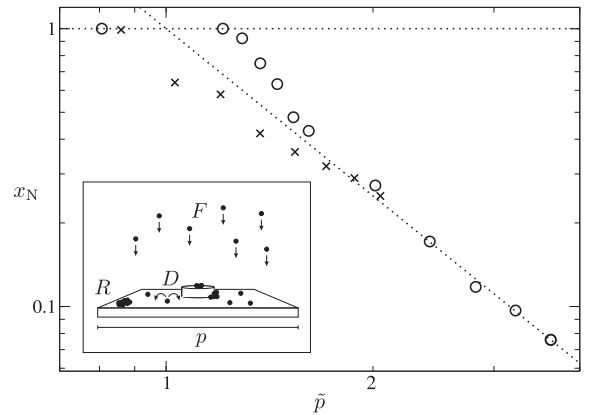}
\caption{\label{FIG:ExpVsSim}Comparison of experimental ($\times$)
and simulated ($\bigcirc$) nucleation control $\xn$ for different
pattern spacings $\dimless{p}$ from Ref.~\cite{ChiPRL}. While the
simulated curve shows a retention of nucleation control
for $\dimless{p}>1$, the experimental curve exhibits an early loss
at $\dimless{p}$ significantly smaller than unity. In the inset
the parameters of the model are defined.}
\end{figure}

From a technical point of view one is interested in exclusive
accretion of the deposited substance at the predefined sites and
no additional nucleation. The predefined sites display a regular
quadratic arrangement (size of unit-cell: $p^2$; see inset of
Fig.~\ref{FIG:ExpVsSim}). To provide a quantitative means of
evaluation, the nucleation control can be expressed by
\begin{equation}\label{EQ:DefXn}
\xn := \frac{1}{1+R}
\,,
\end{equation}
with $R$ representing the number of additional nucleated clusters
per unit-cell. Additionally, to be able to relate theory and
experiment the length scales are renormalized by
\begin{equation}\label{EQ:DefLambda}
\lambda := \sqrt{\left(\frac{A}{N}\right)_\mrm{u}} \quad \text{to} \quad \dimless{p}:=\frac{p}{\lambda}
\,,
\end{equation}
where $(N/A)_\mrm{u}$ represents the overall island density of an
unpatterned substrate. For $\dimless{p} \gg 1$ the influence of
the predefined sites on the nucleation processes can be neglected
so that the nucleation properties just follow the standard
nucleation theory; see,
e.g.,\cite{VenablesPhilosMag,Venables98,Venables00,IslandsMoundsAtoms,EvansThielBarteltReview}.
In contrast, for dense patterns with $\dimless{p} \ll 1$ all
nucleation occurs at the predefined sites ($\xn\approx
1$)~\cite{BardottiNanoclusters,KalischewskiZhuHeuerPRB}.

The transition regime $\dimless{p} \approx 1$ is of particular
interest. First, from an experimental/technological view one needs
to understand how far the range of complete nucleation control
extends to large pattern sizes. Second, from a theoretical
perspective new phenomena come into play as compared to standard
nucleation theory because of the interplay of the nucleation
tendency and the drainage of the adatoms by the predefined sinks.
The complexity of this regime is also highlighted in
Fig.~\ref{FIG:ExpVsSim} where the results of previous computer
simulations~\cite{ChiPRL,KalischewskiZhuHeuerPRB} and experiments
are displayed ~\cite{ChiPRL}. Whereas the simulations display
nucleation control in the regime $\dimless{p} > 1$ the experiments
show a loss of nucleation control at pattern spacings
significantly below $\lambda$ (i.e. $\dimless{p}<1$). What is the interpretation of these differences?

In this paper we present the theoretical basis of nucleation
control  and in particular reveal the influence of the
experimental deposition time scale. For a clear identification of
the physical mechanisms we employ a minimum model which considers
particle deposition at flux $F$ and subsequent surface diffusion
with diffusion constant $D$. Adatoms are either included by the
predefined sites, acting as perfect sinks, or, if the distance
between the sites $p$ is too large or $F/D$ is too small, start to
form additional nuclei ($R$) aside of these locations. Since we
are interested in the key mechanisms, we neglect possible
desorption processes.

To evaluate the temporal evolution of the nucleation behavior on
a pre-patterned substrate a simulation technique similar to the
level-set methods of
Refs.~\cite{PhysRevB.65.195403,PhysRevE.58.R6927,VardavasGrowthWithDefectSites,ChenLevelSetEpitaxial}
was applied. This algorithm is far more efficient as compared to the previous atomistic
Monte Carlo simulations~\cite{ChiPRL,KalischewskiZhuHeuerPRB}.
It is based on the
numerical solution of the adatom diffusion equation
\begin{equation}\label{EQ:DiffusionEquation}
\frac{\partial c(\vect{x})}{\partial t} = D \nabla^2 c(\vect{x}) + F
\end{equation}
by discretization in time and space. The generation of new nuclei follows the concepts of standard nucleation theory~\cite{VenablesPhilosMag,IslandsMoundsAtoms,EvansThielBarteltReview}. The local rate and hence also the chance of nucleation $S_{j,t}$ on the grid point $j$ at time $t$ with the next time step $\Delta t$ is given by
\begin{equation}
S_{j,t} = k (c_{j,t}/c_0)^{\istar+1} \Delta x\,\Delta y\,\Delta t
\,,
\end{equation}
where $\istar+1$ represents the number of adatoms in the first stable two-dimensional nucleus and $k$ and $c_0$ are parameters which characterize the nucleation
probability.  The parameters \cite{data} were adjusted such that we obtained a typical distance of about 50 grid points between nuclei in the steady state regime of the unpatterned case. For the definition of a length scale in our simulations we have introduced the lengthscale $\mrm{u}$, where $\mrm{u}/2$ describes the distance of two adjacent grid points. Eq.~\eqref{EQ:DiffusionEquation} is supplemented by the boundary condition $ c_\mrm{sink} = 0$ at the positions of nuclei.

Commonly (see e.g.~\cite{VardavasGrowthWithDefectSites,PhysRevE.58.R6927,PhysRevB.65.195403}), a new nucleus is generated when $Q(t)$, defined via
$
Q= \sum_{j,t} S_{j,t}$
exceeds an integer value. However, the deterministic nature of the first nucleation event introduces a bias during a critical phase of this simulation~\cite{PhDThesis}. Thus, we apply the following Monte Carlo like algorithm: at every time step, $S_t=\sum_i S_{i,t}$ is spatially integrated and in combination with sufficiently small $\Delta t$ provides $S_t \Delta t\ll 1$. This chance is then compared to a random number $0<r<1$ resulting in nucleation for $r<S_t \Delta t$. In a successful event the new nucleus is placed stochastically according to the individual $S_{i,t}$.

A simulation run consists of one quadratic unit-cell of the surface pattern with length $p$ as shown in the inset of Fig.~\ref{FIG:ExpVsSim}, using periodic boundary conditions. We have checked that the $x_N$ vs. $\dimless{p}$-plot does not display any finite-size effects. The finite-size effects, reported in literature \cite{Wolf95}, occur in a regime where in the present case $x_N=1$. Each data point results from averages over 200 independent simulations.  All systems are subject to the same flux $F$ and a value of $\istar=4$ \cite{KalischewskiZhuHeuerPRB}. Similar to Refs.~\cite{BarteltEvansExactIslandSizeDistribution,PhysRevB.46.12675} we represent the predefined sites as a single grid point on our discretized simulation lattice, acting as a perfect sink. The same holds for newly generated nuclei. In this way a predefined or generated nucleus possesses an effective area $\mrm{u}^2/4$. As we are not interested in the temporal evolution of the individual nuclei, but in the averaged properties of the surface, the sinks remain to be confined to their initial position and do not expand any further.  Experimentally, the area $\Omega$ occupied by an adatom is, of course, much smaller than that of the pattern. Consequently, the ratio $\Omega/\mrm{u}^2$ is orders of magnitude smaller than unity $(\mathcal{O}(10^{-4}))$ and the resulting values for the coverage $\Theta$, which are given below in units of $\Omega/\mrm{u}^2$, hence correspond to sub-monolayer growth.

\begin{figure}[ht]
\includegraphics[width=0.95\linewidth]{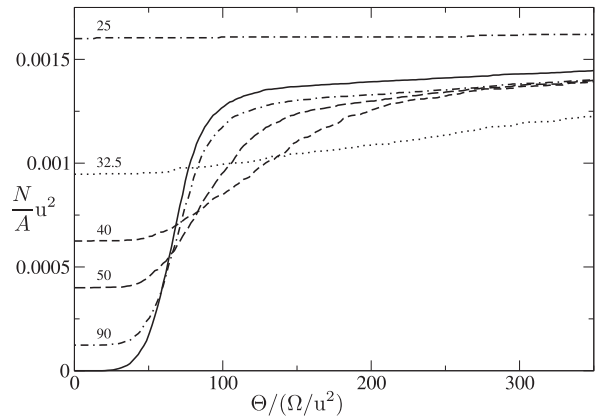}
\caption{\label{FIG:NdensVsTheta}Evolution of the nucleus
densities $N/A$ with growing time.  Shown are surfaces of
different pre-patterned densities (with $p/\mrm{u}$ indicated by
the numbers), as well as the unpatterned reference system (solid
line).}
\end{figure}

For good nucleation control the nucleus density of the
pre-patterned surface $(N/A)_\mrm{pt}$ has to be lower than that
of the unpatterned counter part $(N/A)_\mrm{u}$. In what follows
we compare the temporal evolution of these densities for several
patterned systems of different $p$ complemented by an unpatterned
reference system; see Fig.~\ref{FIG:NdensVsTheta}. The unpatterned
system (solid line) possesses the characteristic shape expected
from standard nucleation theory~\cite{IslandsMoundsAtoms}: The
beginning shows a \emph{transient time regime} of rapid
nucleation,  dominated by local supersaturation. However, as the
surface becomes saturated with nuclei, the overall concentration
decreases and the system enters the regime of \emph{steady state
nucleation}. Mechanistically, the comparatively slow nucleation
rate of this regime can be attributed to statistical fluctuations
in the adatom concentration.

With respect to the patterned surfaces Fig.~\ref{FIG:NdensVsTheta} allows the following observations: (a) For comparably large $p$ (see, e.g, $p/\mrm{u}=90$) the behavior of the patterned systems resembles that of the unpatterned counter part, but as by construction there already exists one dot at the beginning of the simulation, (b) the systems  start off at a density of $(N/A)_\mrm{pt}(t=0)>(N/A)_\mrm{u}(t=0)=0$. As growth commences, however, the permanent adatom depletion by the pattern leads to an effectively smaller nucleation rate such that (c) the patterned surfaces approach the unpatterned long time behavior from below. (d) Furthermore, as $p$ becomes smaller ($p/\mrm{u}=50,40$) the s-shaped part of the curve is stretched in time corresponding to a slower transition through the rapid nucleation regime. This effect increases to the point that (e) at sufficiently low pattern spacing, e.g. $p/\mrm{u}=32.5$, the s-shape becomes unrecognizable and $(N/A)_\mrm{pt}$ remains significantly below $(N/A)_\mrm{u}$ within the considered time scale.  (f) Finally, $p$ can be decreased to the point where $(N/A)_\mrm{pt}(t=0)$ becomes so large that the unpatterned long time behavior is approached from above.

Returning to the question of nucleation control, Eq.~\eqref{EQ:DefXn} can, with the help of Eq.~\eqref{EQ:DefLambda},  be rewritten as
\begin{equation}
\xn = \left(\frac{N}{A}\right)_\mrm{u}\!\!\!(t) \Bigg/ \left(\frac{N}{A}\right)_\mrm{pt}\!\!\!\!(t,\dimless{p}) \;\cdot\; \frac{1}{\dimless{p}^2}
\,.
\end{equation}
A surface can hence only lie above $\xn=\dimless{p}^{-2}$ and show retained nucleation control if $(N/A)_\mrm{u}/(N/A)_\mrm{pt}>1$. Specifically this means that because of (b) the nucleation control at the beginning of the experiment is always very low, but due to (c) and (d) this changes with increasing time for all patterns up to (f).

\begin{figure}[ht]
\includegraphics[width=0.95\linewidth]{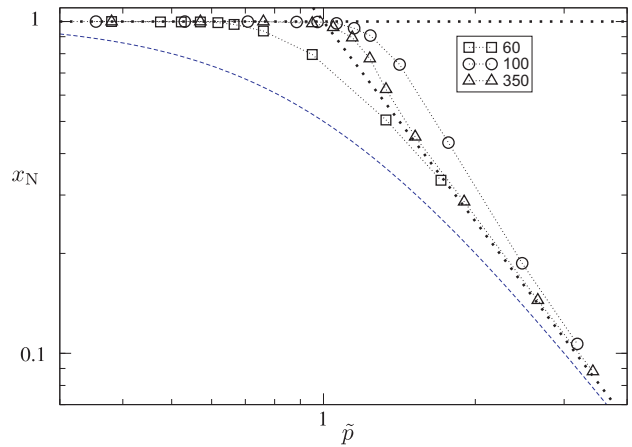}
\caption{\label{FIG:XnVsPatdifTheta}Nucleation control $\xn$ with
increasing coverage $\Theta/(\Omega/\mrm{u}^2)= 60\,(\Box)$,
$100\,(\bigcirc)$ and $350\,(\vartriangle)$. At relatively low
coverage the patterns display an early loss of nucleation control
($\xn$). However, as the unpatterned reference system passes into
the steady state nucleation regime, nucleation control becomes
most pronounced. With further increasing time the curves converge
to the dotted limit. The low-coverage limit $x_N =
1/(1+\tilde{p}^2)$ is indicated by the broken line.}
\end{figure}

The corresponding nucleation control can be found in
Fig.~\ref{FIG:XnVsPatdifTheta}: at low coverage
($\Theta=60\,\Omega/\mrm{u}^2$) the unpatterned surface does not
yet show significant nucleation and hence the vast majority of
patterned systems exhibits a higher nucleus density, which in turn
results in an early loss of nucleation control, indicated by a
curve in the lower left segment of the figure. As time increases,
however, and the regime of rapid nucleation begins around
$\Theta=70\,\Omega/\mrm{u}^2$, the unpatterned nucleation density
commences to exceed its patterned counterparts leading to an
increasing number of systems under nucleation control and hence to
a crossing of the $\xn=\dimless{p}^{-2}$ diagonal closer to
$\dimless{p}=1$. As the unpatterned surface reaches the beginning
of the steady state nucleation regime, nucleation control becomes
most pronounced which can be seen from the curve corresponding to
$\Theta=100\,\Omega/\mrm{u}^2$. Past this point the slow
convergence of all patterned nucleus densities to that of the
unpatterned surface results again in a decrease of nucleation
control.  All this effects together mean that at fixed $\tilde{p}$
one has a non-monotonous dependence of $x_N$ on $\Theta$. Note
that upon changing the coverage the onset of nucleation control
can vary as much as a factor of two with respect to the critical
pattern density (relative to the respective value of $\lambda$).

Two limits can be treated more quantitatively. First, in the limit
of low coverage and large $p$ ($\Theta p^{2} = const$) the growth
of additional nuclei is not influenced by the gold dot. Thus one
has $(N/A)_\mrm{pt} = (N/A)_\mrm{u} + p^{-2}$, which yields $x_N =
1/(1 + \tilde{p}^2)$. This limit is included in
Fig.~\ref{FIG:XnVsPatdifTheta}.

Second, further analysis is possible when the first nucleus is
formed from a stationary concentration profile. As shown in
Ref.~\cite{KalischewskiZhuHeuerPRB} for fixed parameters one can
find a critical time scale $t_c$ such that for $t > t_c$ the
stationary concentration profile is reached. For this condition to
be applicable it is important that nucleation can be neglected for
times smaller than $t$ or, equivalently, $R < 1$ at time $t$. The
time scale $t_c$ scales like $p^2$ \cite{KalischewskiZhuHeuerPRB}.
In this scenario one can derive the relation
\begin{equation}\label{EQ:ResultR}
R\propto \Theta \dimless{p}^f
\end{equation}
where $f$ can be calculated from  knowledge of the stationary
concentration field, which was done numerically in
Ref.~\cite{KalischewskiZhuHeuerPRB}.

\begin{figure}[ht]
\includegraphics[width=0.95\linewidth]{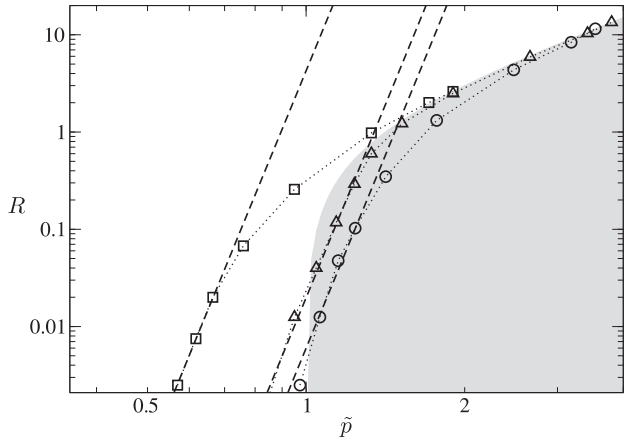}
\caption{\label{FIG:RVsPatdifTheta}Additionally formed nuclei $R(\dimless{p})$ at increasing coverage $\Theta/(\Omega/\mrm{u}^2)= 60\,(\Box)$, $100\,(\bigcirc)$ and $350\,(\vartriangle)$.  The theoretical expectation for the static case according to Ref.~\cite{KalischewskiZhuHeuerPRB} ($R\propto \dimless{p}^f$ with $f=13.2$) is represented by the dashed lines. The regime of retained nucleation control is indicated in grey.}
\end{figure}

In Figure~\ref{FIG:RVsPatdifTheta} we have replotted
$x_N(\tilde{p})$ as $R(\tilde{p})$. In this representation one can
clearly see that Eq.\ref{EQ:ResultR} is recovered for small values
of $\dimless{p}$. This is to be expected because for a given time
scale of the experiment the stationarity condition is always
fulfilled in the limit of small $p$ and correspondingly small
$t_c$. Furthermore the linearity with respect to $\Theta$ is
reflected in the linear increase of $(N/A)$ for small $p$ in
Fig.~\ref{FIG:NdensVsTheta}. Interestingly, the applicability
range of this power law strongly depends on the coverage: while at
$\Theta=60\,\Omega/\mrm{u}^2$ the relation only holds for values
less than $R \approx 0.05$, for larger coverage the power law
nearly holds up to the theoretical limit of $R=1$. This can be
easily understood from analyzing $R$ at the time scale $t_c$,
introduced above. Joining together Eq.~\eqref{EQ:ResultR} and the
relation $t_c \propto p^2$ one obtains $R(t_c) \propto t_c^{f/2}$.
Indeed, one observes a dramatic increase of the range of
applicability of Eq.~\eqref{EQ:ResultR} upon increasing the
coverage. Of course, at the latest the power law scaling has to
break down when $R$ approaches 1.

Based on these considerations we can conclude the following: the experimentally observed early loss of nucleation control as depicted in Fig.~\ref{FIG:ExpVsSim} does not disagree with the general understanding of the mechanisms governing template-directed growth, and can be attributed to dynamic effects. In particular, this effect is caused by a different temporal evolution of the nucleus density on the patterned and unpatterned substrates: unpatterned substrates or patterns of very large $p$ show a distinct separation between the transient and the steady state regime, which mechanistically corresponds to nucleation by supersaturation or by statistical aggregation, respectively. With decreasing pattern spacing, however, the transient regime becomes dilated until it is phenomenologically as well as mechanistically indiscernible from steady state nucleation. In the commonly used representation, which is normalized by an unpatterned surface, this leads to the effect that initially the majority of patterned systems exhibits an early loss of nucleation control. This changes, however, as the unpatterned surface approaches the end of the transient regime which leads to maximal nucleation control close to the beginning of steady state nucleation. Thus, a consistent picture of the influence of the distance of the patterns as well as the experimental time scale on nucleation control can be formulated. This is of particular relevance for the understanding of the formation of nanostructure on surfaces. Evidently, also more complex situations such as non-circular patterns can be analyzed within the present model approach.

We acknowledge very helpful discussions with L.F. Chi and her group as well as with S. Hopp and T. Mues. F. K. thanks the Fonds der Chemischen Industrie and the NRW Graduate School of Chemistry for financial support.


\begin{thebibliography}{23}
\bibitem[{dat()}]{data}
\bibinfo{note}{We have used the parameters $k = 10^{-9}/(u^2s)$, $c_0 = 10 \,
  {\rm particles}/u^2, F = 0.0075 \, {\rm particles}/(u^2s)$ and $D = 2u^2/s$.}
\end{thebibliography}

\end{document}